# Dual State-Space Model of Market Liquidity: The Chinese Experience 2009-2010.

Peter B. Lerner[1]

**Abstract**

This paper proposes and motivates a dynamical model of the Chinese stock market based on a linear regression in a *dual* state space connected to the original state space of correlations between the volume-at-price buckets by a Fourier transform. We apply our model to the price migration of executed orders by the Chinese brokerages in 2009-2010.

We use our brokerage tapes to conduct a natural experiment assuming that tapes correspond to randomly assigned, informed and uninformed traders. Our analysis demonstrates that customers' orders were tightly correlated—in a highly nonlinear sense of neural networks—with the Chinese market sentiment, significantly correlated with the returns of the Chinese stock market and exhibited no correlations with the yield on bellwether bond of the Bank of China. We did not notice any spike of illiquidity transmitting from the US Flash Crash in May 2010 to trading in China.

.

## 1. Introduction

The technical advantage of speed was always exploited by the traders. While the case of London's Rothschild receiving detailed information about Napoleon's movements on the continent can be anecdotal, the use of postal pigeons existed since antiquity and Reuters (2008) documented their own service across the English Channel since XIX century. The first electromechanical fax communicated stock quotes between Lyon and Paris in the 1860s installed by physicist and a priest Caselli (1865) but it was highly impractical because of contemporary limitations on technology and was soon replaced by sending coded messages through telegraph. The arms race for the execution time continues to this day.

Order execution became fast—time stamps on the order of minutes and seconds were common in early XXI century—but now have reached the microseconds, for which relativistic limitations on signal transmission became essential (Angel, 2014). To keep up with the progress one has to find a method of analysis, which is largely independent on the extant technology, i.e. practically, of the market latency and trading algorithms.

---

[1] Kean State University, Wenzhou campus (retired). I thank my former student, Lin Jilei (U. Illinois) for invaluable help.

There are other complexities in the empirical market microstructure apart from execution speed, incomplete or frivolously manipulated data. One of them is that order flow "lives" in transaction time rather than in physical time (Hasbrouck, 2016, Xiaozhou, 2019), another that the real trading costs can be hard to estimate and relatively easy to conceal (Keim, 1998). We partially circumvent limitations of the first kind by using interday correlations of intraday price migration through volume buckets as regression panels (see Section 3 below). Interday correlations should be free from the absence of transaction time stamps because of the T+1 rule, unique to the Chinese stock markets described by Ming Guo et al. (2012), and Qiao and Dam (2020).

Many semi-empirical measures have been used to describe market behavior on a microstructure level (Fong, 2017). The most-popular, or well-researched theoretically are VPIN (Easely, 2012), volume imbalances (Humphery-Jenner 2011, Lipton 2013, Cartea 2015a), VWAP and its modifications (Cont 2014, Cartea2015b, Cartea2015c, see also Lerner, 2015) and all the different versions of Amihud measure (Amihud, 1986, Hasbrouck, 2007).

The application of the state space approach to the analysis of microstructure is not new. Hendershott and Mankveld (2014) specifically emphasized this line of research with respect to HFT as an alternative to the autoregressive model. A standard way to analyze state-space distributions is Kalman filtering and, in particular, Rzayev and Ibikunle (2019) implemented it to distinguish between liquidity-driven and informed trading component of the trading volume for S&P500. We absconded Kalman filtering because it seemed to drastically reduce variability of our dataset.

This paper continues previous line of research by one of the current authors (Lerner, 2018). It suggests a study of the state space *dual* to the price-to-volume distribution and connected with the state space by Fourier transform. We employ a correlation measure on the state space invented by Roll (1984) but here applied to the price bucket in its entirety rather than to individual stocks. Taking correlations as a first stage of data de-noising has been used by Lehalle (2013), in particular, for his studies of the "Flash Crash" of May 6, 2010.

Three types of volume per weighted price distributions were analyzed—the "Buy" volumes, the "Sell" volumes and the imbalances volumes, which could be half-seriously called BVWAP, SVWAP and IVWAP, though we abstain from this terminology in the rest of the paper. Our definition of imbalance slightly varies from Cont, 2014, where it is defined as twice the difference between buy and sell volume divided by their arithmetic mean. Effectively, we use geometric means (see Section 3).

In the case of imbalances, our measure is similar to the VPIN distribution proposed by Easley, Prado and O'Hara (Easley, 2012), except that it does not involve computation of intra-bucket price variance. Instead we use day-to-day correlations of volumes within a given price bucket. We also employ Amihud-like measure (Amihud, 1986) to test the contagion between America and China during the days surrounding the Flash Crash.

To test our methodology we used brokerage tapes of a number of Chinese brokerages submitted to mainland Chinese stock exchange as a matter of regulatory compliance during 2009-2010. These tapes contain only completed trades. They do not have time stamps beyond one day but display most of the trades with "Buy" or "Sell" indicators across the entire price range. Because the tapes divide trades by "Buy" and "Sell" (less than 10% of the records miss this stamp), we do not rely on algorithmic estimation of this division as in Lee and Ready (1991). Whatever incompleteness exist in our data, it lies in reporting procedures for the brokerages, which existed during these years.

To analyze the volumetric data, we combine them into uniform buckets of 0.5 CNY so that a typical number of buckets is around 150-180 at any given day during 2009-2010. Then we build a dual space model of trading, which we further analyze by the learning algorithms.[2]

In that analysis, we follow in the footsteps of Foster and Viswanathan (1996) who developed a theoretical model of several group of traders who try to predict actions of others. Our model allows to gauge how these predictions could have panned out empirically. We use three metrics of market reaction: the Chinese market sentiment (Baker and Wurgler, 2006, Hu, 2012, Liu and An, 2018), returns on the Shanghai stock market and yields on a Bank of China 10-year bond. Using our model we can directly, and relatively parsimoniously, explore the conditions prevailing in dark pools, artificially supplying or denying our assumed traders external information about the activities of their colleagues or direction of the indexes.

---

[2] Dual, and connected by the Fourier transform with respect to the chosen trade measure on the state space.

The paper is structured as follows. In the next section, we provide extensive statistics of databases at our disposal. In the Section 3, we describe the state space of the problem. In the Section 4, we build the model of predictable trade and describe its inputs and outputs. In the Section 5, we provide validation for our model for predictable variation of trading intensity. The residuals of our prediction model are being analyzed through shallow and deep learning networks imitating the decision process of the traders in response to the new events. We discuss information, which can be gleaned by the fictitious traders in the subsequent Sections 6. Finally, in the Section 7, we produce a single event study. The paper is completed by the Conclusion.

## 2. Summary statistics of the databases

Brokerage tapes provided as Excel files have the following format shown on the Fig. 1.

| Trddt       | Stkprc        | Parcha     | Trdtims   |
|-------------|---------------|------------|-----------|
| Trading Da  | Trading I     | Nature O   | Number Of |
|             | CNY           |            | Deal      |
| 2009-08-06  | 10.05         | S          | 425       |
| 2009-08-06  | 10.2          | S          | 81        |
| 2009-08-06  | 10.17         | S          | 321       |
| 2009-08-06  | 10.01         | B          | 451       |
| 2009-08-06  | 9.95          | B          | 393       |
| 2009-08-06  | 9.99          | B          | 30        |
| 2009-08-07  | 9.9           | B          | 708       |
| 2009-08-07  | 9.75          | B          | 460       |
| 2009-08-07  | 9.7           | B          | 608       |
| 2009-08-07  | 9.7           | B          | 444       |
| 2009-08-07  | 9.6           | B          | 299       |
| 2009-08-07  | 9.83          | S          | 131       |
| 2009-08-07  | 9.61          | B          | 192       |

Fig. 1 Printout from one of the brokers' tapes.

They include date, order price in CNY, type of order "buy" and "sell", as well as the order's volume. Further on, we classify separate spreadsheets as "tapes" from zero to four for all brokerages though, more likely, these were just arbitrary partitions of the order book into spreadsheets. As we can see from the data in the Table 1, extensive statistics for individual traders is comparable and we treat them as an extra level of randomization of our data.

The tapes do not distinguish between individual stocks. We can only surmise that the trades within a given price bucket belong to a single, or maximum two stocks given a comparatively thin volume of trading in the stock markets of Mainland China during 2009-2010.

Our database included eight brokerages with the symbols *rfokp4c*, *hvw5se4*, *zbe0rgv0*, *qwixupca*, *qguyi05q*, *gxbmxv0*, *q1ysmbyz* and *5vuyp3bu*. Only three last of these brokerage tapes contained complete data on the volume, which we further denote by the first symbol as "g", "q" and "5". Most numerical examples in this paper as well as the data refer to the Brokerage 5.

Despite the fact that, in most likelihood, the division of trades between tapes is arbitrary, for the purpose of later analysis we shall imagine them as belonging to separate "traders". This corresponds to the intuitive idea that in the modern high-frequency trading, the trader is a computer algorithm, which arbitrarily parses the state space.

Our tapes contain (see Table 1), several tens of thousands trades for the period of two years. If one wants to project this rate on the intensity of the modern high-frequency trading, all the tapes of one brokerage would correspond to 7-8 minutes of fast trading.[3] This illustrates utility of observing emerging markets where the phenomena requiring grotesque amount of data to analyze, can be observed with much less granularity.

---

[3] Characteristic time of the first response on a trading signal is $\tau \approx$ 2-3 msec., which roughly corresponds to the computer messages cycling the circumference of New York City and vicinity with the speed of light, Hasbrouck, 2016. Inherent latency of trading quotes is even shorter, see Bartlett and McCrary (2019), Table 1.

Table 1. Summary statistics of the Brokerage 5 for the 484 trading days between the beginning of 2010 and the end of 2011. Traders are marked as 0 through 4. Volume data are rounded to one share. The stock price is expressed in Chinese Yuan. The index B or S refers to "Buy" or "Sell" order on a given tape. We used sample volume variance as an indicator of volume dispersion. Daily standard deviation of the volume can be approximated as $\sqrt{484}$ X $\sqrt{SVV}$. There is no systematic difference between tapes from 0 to 3, while records in the tape 4 gravitate towards higher-priced stocks.

| Tape /Type order | No. of Trades | Min Price, CNY | Avg. Price, CNY | Max Price, CNY | Std. Price, CNY | Avg. Daily Volume | Sample Volume Variance |
|---|---|---|---|---|---|---|---|
| 05B | 32041 | 2.34 | 8.77 | 63 | 6.17 | 11,046 | 37,535 |
| 05S | 28513 | 2.36 | 8.63 | 61.69 | 5.83 | 9,691 | 33,411 |
| 15B | 31550 | 2.3 | 9.91 | 74 | 5.94 | 10,792 | 33,999 |
| 15S | 28864 | 2.23 | 9.52 | 77.95 | 5.48 | 9,963 | 29,251 |
| 25B | 31988 | 2.57 | 9.84 | 96 | 6.29 | 9,509 | 22,946 |
| 25S | 28584 | 2.53 | 9.58 | 95.1 | 6.09 | 8,468 | 18,797 |
| 35B | 31824 | 2.29 | 12.36 | 69.91 | 6.35 | 9,144 | 29,004 |
| 35S | 26867 | 2.31 | 11.96 | 67.69 | 6.13 | 7,632 | 26,000 |
| 45B | 19975 | 3.95 | 16.95 | 123.5 | 9.24 | 4,939 | 38,321 |
| 45S | 15095 | 3.9 | 9.34 | 193.8 | 9.34? | 3,906 | 65,500 |

## 3. Description of the model

We apply three stages of data analysis in our model. In the first stage, we allocate all daily orders to the price buckets, or buckets. From these price buckets we construct a state space from the day-to-day correlations of order volumes, which we use as new vectors of our state space.

$$\vec{X}_t(x) = Corr(V_{t+1}, V_t)|_{x=price\_bucket} \qquad (1)$$

The Equation (1) was written in an assumption that day-to-day correlations of order volumes exhibit more stability then the volumes itself. Heuristically, this assumption is supported by the unique T+1 rule existing in the Chinese stock markets—one has to hold stock one day or more before selling (Ming Guo, 2012), so that intraday noise must be uncorrelated between today and tomorrow.

Mathematically, one can make a following observation about the correlations. Let $\tilde{u}_t = u_t + \epsilon_t^1$ and $\tilde{v}_t = v_t + \epsilon_t^2$ be our time series, where $u_t$ and $v_t$ are signals with the correlation coefficient ρ and epsilon terms are microstructure noises with variances $\sigma_{\epsilon 1}^2$ and $\sigma_{\epsilon 2}^2$ uncorrelated with the signals and each other: $E[u_t, \epsilon_t^1] = E[u_t, \epsilon_t^2] = E[v_t, \epsilon_t^1] = E[v_t, \epsilon_t^2] = E[\epsilon_t^1, \epsilon_t^2] = 0$. Then, their noisy correlation for a small white noise becomes:

$$\hat{\rho} = Corr[\tilde{u}_t, \tilde{v}_t] \equiv \frac{Cov[\tilde{u}_t, \tilde{v}_t]}{\sqrt{Var[\tilde{u}_t]}\sqrt{Var[\tilde{v}_t]}} = \frac{Cov[u_t, v_t]}{\sqrt{Var[u_t]+E[(\epsilon_t^1)^2]}\sqrt{Var[v_t]+E[(\epsilon_t^2)^2]}} \approx \frac{Cov[u_t, v_t]}{\sqrt{Var[u_t]}\sqrt{Var[v_t]}}\left(1 - \frac{1}{2}\left(\frac{\sigma_{\epsilon 1}^2}{Var[u_t]} + \frac{\sigma_{\epsilon 2}^2}{Var[v_t]}\right)\right) \leq \rho \quad (2)$$

This is a downward biased estimator of a true correlation, stationary, as long as the second moments of the noise are small and stationary. Furthermore, the correlation coefficients are concentrated in the [-1, 1] range and, thus, are more amenable to intuitive and graphic interpretation.

Covariance matrices could be more consistent from the mathematical point of view but they are harder to interpret intuitively—in particular, because they grow as squares of volume with more active trading—and visualize. Moreover, covariance matrices because of their nonlinear growth with average volume obscure participation of the "high impact trades", i.e. the trades, which influence price in excess of their size (Xiaozhou, 2019). As "volumes" in Equation (1) we use three types of variables—volume of buy trades, volume of sell trades and imbalance volume, which we define as difference between buy and sell volume in a given price bucket. In the case of imbalance statistics, our measure is pretty similar to VPIN proposed by Easley, Prado and O'Hara (Easley, 2012).

Our state space is a discrete space of sixteen price buckets, separated by Δ=0.5 CNY (daily changes of stock prices by more than 16Δ=8 CNY were almost never observed in the sample).[4] A smaller granularity will leave too few events in each bucket to allow confident averaging, larger granularity will average over most daily price changes. The division of trading book into equal buckets allows us to avoid a problem that the trades in our database are not stamped with the name of a particular stock. Trading volume of all stocks experiencing "zero" or "significant" price change goes into the same bucket. Our construction of the phase space potentially allows two ways of analysis: panel analysis based on individual buckets and time-series analysis, which follows evolution of buckets through time.

---

[4] We split all the trading book into 0.5 CNY buckets by the price change (a few hundred encompassing all the stock price range) but use only the first sixteen buckets. Using a larger number typically causes a spurious periodicity in our data.

Second stage of our analysis is building a dynamic model of trading. This model is described below. Our only presumption is that dependent and independent variables are being connected by a linear Hilbert-Schmidt (Danford, 1963, Gohberg, 1977) operator:

$$\vec{X}_{t+1}(x) = \int K(x - x') \cdot \vec{X_t}(x') d^n x' \tag{3}$$

The form of kernel in the equation (3) for a discrete state space is a matrix, which we estimate from our data. For our analysis, we use a dual state space obtained by Fourier transform of the initial state space of a model. Philosophically, our choice of the state variable is based on Bochner theorem in functional analysis stating that covariance of a weakly stationary always has a representation as a Fourier integral of a stationary measure (e.g. Stein, 1999). Henceforth, a broad class of stochastic processes can be represented in the form below (Liptser and Shiryaev, 1978, Chapters 14 and 15). Here, we only display our model in the form we had used in our analysis.

Fourier transform of the Equation (2) gives a linear regression in a dual state space:

$$\Delta \vec{X}_{t+1,\omega'} \equiv \vec{X}_{t+1,\omega'} - \vec{X}_{t,\omega'} = \hat{\beta}_{\omega'\omega} \vec{X}_{t,\omega} + \vec{e}_{t+1,\omega} \delta_{\omega'\omega} \tag{4}$$

In the Equation (3) because of the Fourier transform, the vectors are assumed to be complex, i.e. with twice the dimensionality of an original state space. Heretofore, the beta matrix has dimension 32×32 if we separate real and complex parts. Kronecker delta in the regression residual assumes that all spurious correlations between volumes disappear in one-two days.

Note, that we make no assumption about the random process governing the price dynamics. The only limitation of Equation (4) is the size of beta matrix we use to approximate a continuous pseudodifferential operator (Taylor, 1981, Hörmander, 1987), see Appendix. Inverse Fourier transform of our beta operator is analogous to *Q* operator in Huang, Lehalle and Rosenbaum (2014) Markovian model of Limit Order Book (LOB).

Of course, because our initial state vectors are real, there is covert symmetry in the coefficients and a number of rows in the beta matrix are identical zeroes. Original daily state vectors are recovered through the inverse Fourier transform below where the hat denotes the predicted independent variable. They can have a small complex part because of finite representation of decimals in the computer, which we ignore.

$$\vec{X}_t(x), \widehat{\vec{X}_t}(x) = F^{-1}_{\omega \to x}[\vec{X}_{t,\omega}, \widehat{\vec{X}}_{t,\omega}] \tag{5}$$

And

$$\vec{\varepsilon}_t(x) = F^{-1}_{\omega \to x}[\vec{e}_{t,\omega}] \tag{6}$$

At the third stage of our analysis, we employ neural networks to make sense of the regression residuals—whether they are reflecting real economic surprises or a result of noise trading. Because we do not know prediction algorithms used by the traders and, with time, they might become more complicated than anything we can come up with, we try (not very) deep learning backwards. Namely, given an unpredictable part of the day-to-day volume correlations, we try to predict realized indexes of the Chinese economy. The intuition behind this method is that if there is systematic unexpected buying or selling pressure in the market, it must reflect a prevailing market sentiment.

4**. First validation of the model**

We have tested our model's beta estimator for different traders in our database. Our results are being represented by the sets of 484×16 matrices (the number of trading days during 2009-2010 times the number of the price buckets). The correlations between columns and rows of the matrix $\hat{\beta}_{\omega\omega'}$ for the imbalance volumes in the formula (3) are given in the Table 2. Complex beta matrices have dimension 32×32 because of the real and imaginary part of Fourier-transformed state vectors. Yet, under inverse Fourier transform, because of the internal symmetry, both the prediction and residual vectors are real.

We display temperature maps of an estimation of a single tape in the Fig. 2

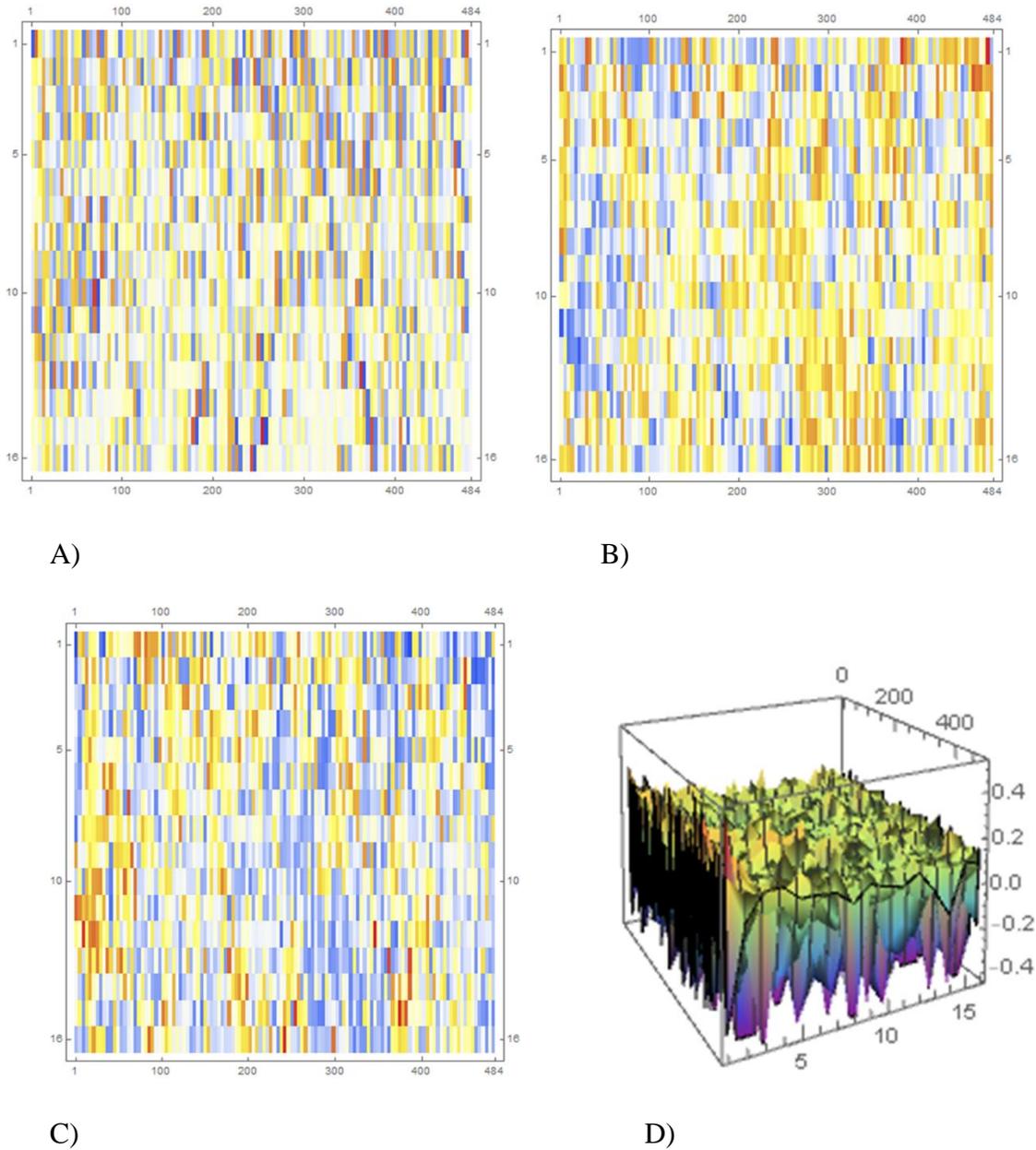

A)

B)

C)

D)

Fig.1. Application of the model of state space regressions (Equations (6-7)) to one brokerage tape. Horizontal coordinate is the observation day and the vertical coordinate is the price change bucket from 0 to 8 CNY in units of 0.5 CNY. A) Temperature map of an original signal. Orange-yellow colors describe a positive day-to-day correlation, light blue and dark blue color—a negative correlation. B), D) predictions of the model, C)—model residuals.

Table 2. Correlations in the complex beta matrices of the linear regression (4) between columns and rows for the correlation of imbalance volumes (see Equation (1)) between the data in different tapes in the Table 1. Correlations in matrix coefficients are symmetric across the diagonal. All the correlations between coefficients are insignificantly different from unity.

A) Correlation between columns of beta matrix of coefficients for different traders

|        | Tape 0 | Tape 1 | Tape 2 | Tape 3 | Tape 4 |
|--------|--------|--------|--------|--------|--------|
| Tape 0 | 1      | 0.9620 | 0.9630 | 0.9635 | 0.9633 |
| Tape 1 |        | 1      | 0.9618 | 0.9563 | 0.9606 |
| Tape 2 |        |        | 1      | 0.9601 | 0.9678 |
| Tape 3 |        |        |        | 1      | 0.9618 |
| Tape 4 |        |        |        |        | 1      |

B) Correlation between rows of beta matrix of coefficients for different tapes

|        | Tape 0 | Tape 1 | Tape 2 | Tape 3 | Tape 4 |
|--------|--------|--------|--------|--------|--------|
| Tape 0 | 1      | 0.9416 | 0.9424 | 0.9574 | 0.9431 |
| Tape 1 |        | 1      | 0.9440 | 0.9421 | 0.9479 |
| Tape 2 |        |        | 1      | 0.9635 | 0.9700 |
| Tape 3 |        |        |        | 1      | 0.9452 |
| Tape 4 |        |        |        |        | 1      |

The fact that in testing regression (3) for the five data tapes, beta matrices are practically identical, though the state vectors are vastly different, suggests robustness of our model for the predictable component of the daily correlation of the imbalances. A similar picture was observed from correlating betas between Buy and Sell tapes as well.

## 5. Predictable component of the bid-ask volume correlations

As a criterion for the quality of approximation of the Equations (5) and (6), we use the vector error estimator for the predictor and the residuals:

$$F_n = \frac{1/T \sum_{t=0}^{T} \vec{\varepsilon}_t^{\,2}(x_n)}{1/T \sum_{t=0}^{T} \vec{X}_t^{\,2}(x_n)}$$

$$P_n = \frac{1/T \sum_{t=0}^{T} \vec{\tilde{X}}_t^{\,2}(x_n)}{1/T \sum_{t=0}^{T} \vec{X}_t^{\,2}(x_n)} \equiv 1 - F_n \tag{7}$$

In Equation (7), we retained multiplier and *1/T, T=484* (trading days) in denominator as well as numerator for clarity. Index *n=1÷16* numbers a vector of the state space. A typical plot of variances of the predictor and the residual, is given at the Fig. 1:

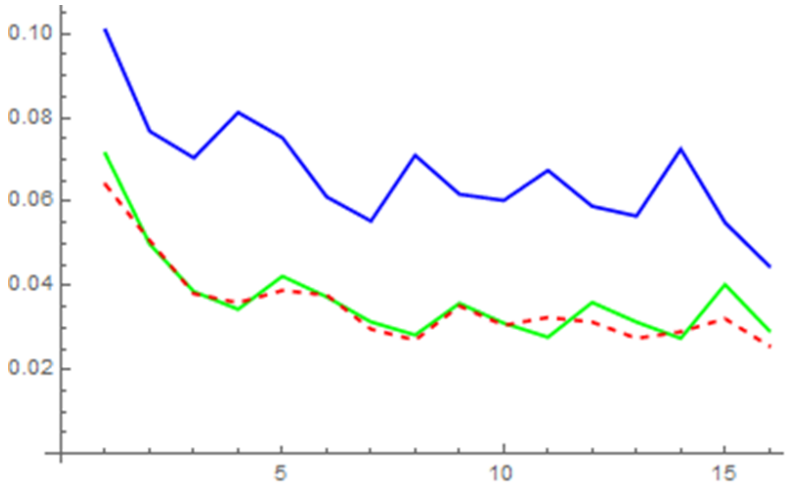

Fig. 2. Respective variances of the state vectors of correlations, n=1÷16 for the trade imbalances for one, randomly chosen, tape. Blue line signifies empirical data from one of the tapes, green line—the predictor variance $P_n$, and the red dash—the residuals variance $F_n$ both integrated for 484 trading days. We observe, that the variance of the empirical distributions is being split approximately 50:50 between the predictor and the residual.

We note that the predictor and the residual time series by construction have zero correlation. Yet, the coincidence of time-weighted variances between the price buckets in Fig. 1 is quite impressive and it is typical for tally of the imbalances.

For quantifying the determination of regression prediction and regression residuals, we used running correlations of panel variances for the 484 trading days in the sample.[5] Matrix of these correlations are provided in the Table 2. From this matrix we observe that 30-40% of the daily variability of the traders' samples and 50% of the monthly variability is contributed by the prediction variance, and the rest—by the regression residuals. Correlations between different brokerage tapes are statistically insignificant. This exercise suggests that the trader's samples are independent in the sense of linear regression. For the trader it means that processing data from another trader by the linear regression does not contribute any valuable information.

Table 3. Matrix of determination ($||\rho_{ij}||=||r_{ij}^2||$) for the variances of regression predictions and regression residuals. Matrix elements are the squares of $r_{ij} = Correlation_t[Var_\Delta(\hat{Y}_i), Var_\Delta(\varepsilon_j)]$ where $i=1 \div 5$ and $j=1 \div 5$ are individual traders. Individual traders can fairly predict their own correlations between today and next day volumes, i.e. persistence of their own demand across all price buckets, but not correlations for other traders.

| Traders' # prediction/residuals | 1 | 2 | 3 | 4 | 5 |
|---|---|---|---|---|---|
| 1 | 0.4126 | 0.0050 | 0.0003 | 0.0002 | 0.0006 |
| 2 | 0.0001 | 0.3081 | 0.0410 | 0.0052 | 0.0052 |
| 3 | 0.0003 | 0.0219 | 0.3639 | 0.0390 | 0.0374 |
| 4 | 0.0007 | 0.0039 | 0.0163 | 0.3408 | 0.0364 |
| 5 | 0.0001 | 0.0010 | 0.0219 | 0.0140 | 0.4027 |

---

[5] Remember that correlation of predictions and residuals is machine zero by construction.

## 6. Analysis of the phase space regression residuals

The model of Equations (3-4) describes a predictable component in the day-to-day correlations of trading volume of price migrations including the zeroth bucket (price changes below 0.5 CNY). Residuals contain both microstructure noise and reaction to unpredictable economic events in the market. To analyze the residuals, we employ several methods inspired by the neural networks.

Because we do not know what kind of training algorithms traders might be using and, given a quick progress in the algorithmic finance and computation power since 2009-2010 and even as this paper is being written, we employ the following method. Instead of prediction of the trading data, we try different simple methods of backdating market data, namely, Chinese market sentiment index, returns on Shanghai stock index and yields on the bellwether 10-year bond of Bank of China. Because of monthly periodicity of the sentiment index, for the consistency of our tests, we used monthly stock returns and monthly bond yield as well.

We certainly cannot match sophistication of algorithms being used by the modern HFT firms and hedge funds and the computational power available to them (Davey, 2014, Yang, 2015, Lopez de Prado, 2018), though our "primitive" algorithms could have been closer to the state-of-the-art in 2009-2010.

In our case, real trading algorithms would have to predict the "unexpected" direction of price changes imprinted in brokerage orders given their information on the markets. Yet, we assign to our imagined traders—represented by the brokerage tapes—a much simpler task of predicting a monthly index given their observation of day-on-day correlation of orders within a given range of price change.[6]

Our procedure corresponds to the following stylized situation. We select a randomly chosen "informed" trader, who observes orders from her own clients and trains her network by predicting the index. We use her data as a network input and then simulate the behavior of other traders whom we consider uninformed as to the direction of the three chosen indexes but who controllably can observe or be in the dark concerning the actions of their colleague from the same brokerage (Fig. 3).

---

[6] This reasoning is based on unproven but intuitive assumption that an economically simpler problem—guessing an "covert" index from proprietary trading data—rather than other way around, is algorithmically simpler.

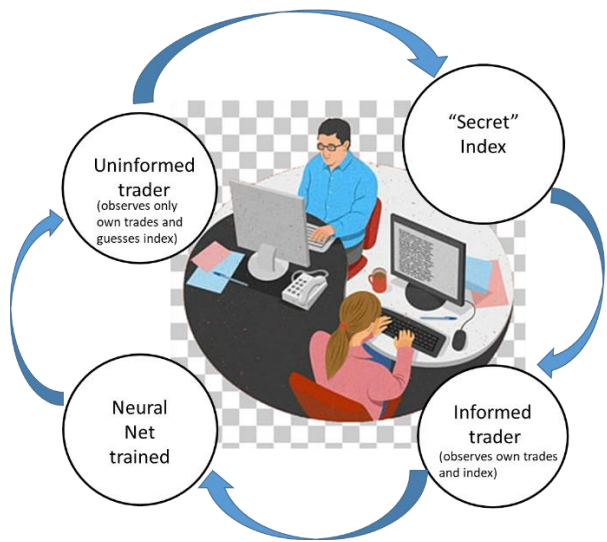

Fig. 3 The protocol of our tests. "Informed" trader observes an index and her own orders and trains the net. "Uninformed" trader uses his own orders and a trained net to backtest the index.

The situation of "leaky brokers" has been described in Maggio (2019) in the following terms: "When considering the theoretical soundness of a market equilibrium in which brokers leak order flow information, one may wonder why an informed asset manager is willing to trade with brokers that tend to leak to other market participants… The broker would enforce this cooperative equilibrium across subsequent rounds of trading. In particular, the broker can exclude from the club the managers that never share their private information and reward with more tips the managers that are more willing to share".

The first test used prediction of three indexes from the first four average moments using a shallow neural network with one hidden layer. When we feed the network data from the stock imbalances, shallow learning network predicts a constant answer indicating zero information about direction of the indexes. This negative result is not all bad because it suggests that the distribution of the residuals is practically indistinguishable from a normal distribution—a nice enough feature for such an artificial model.

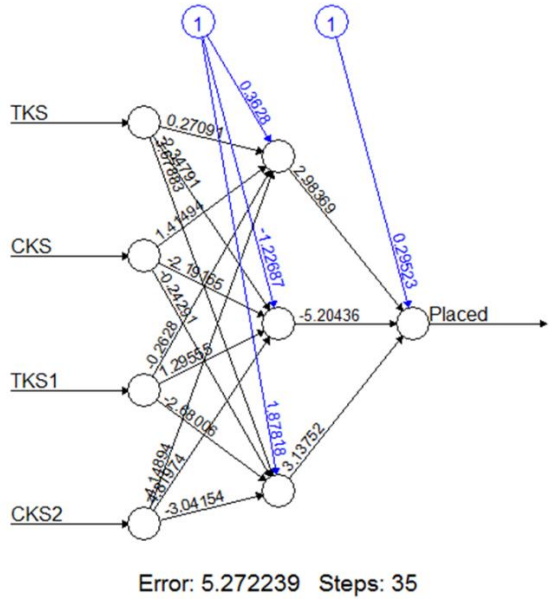

Fig. 4. Example of estimating of shallow three-layer network with one hidden layer for the prediction of the market sentiment (stock returns, bond yields) from the first four moments of the error distribution. The input data for the network are pre-processed and table of the first four monthly moments are being fed into the network. Blue arrows from circles marked as '1' signify training of one sample by the actual market index data.

For our second test, we have used a complete set of residual matrices and the following experimental procedure (Fig. 4). We trained a relatively deep 10-layer network on a sample of all monthly regression errors from a randomly chosen trader omitting the last day of each month and tested this prediction on the same trader's data sample of last days of the month. Then we tried to predict our indexes (sentiment, stock returns and bond yields) back from the supposedly unexpected changes—provided by our regression―in other trading tapes.

One of the difficulties in dealing with neural networks is that results frequently represent multidimensional tensor. By their origin, they cannot be listed compactly in two-dimensional tables and their presentation on a sheet of paper or a computer screen is necessarily confusing to the human perception. We shall return to this difficulty in the next section.

We present the results of 10-layer network in the Table 4. Training on buy or sell signals had some limited power for predicting market indexes from a sample of imbalance correlations for the same trader, or imbalances for other traders whom we considered "blind", i.e. who predict direction of indexes from their own imbalance quotes. Some outputs from this model are shown in Fig. 5.

The inverse path using the same network, i.e. predicting indexes by the training sets on unexpected change in imbalances was added for control. The scalar network results show little dependence on the number of training rounds and functional shape of individual transmission function (ReLu—rectified linear unit, hyperbolic tangent or logit).

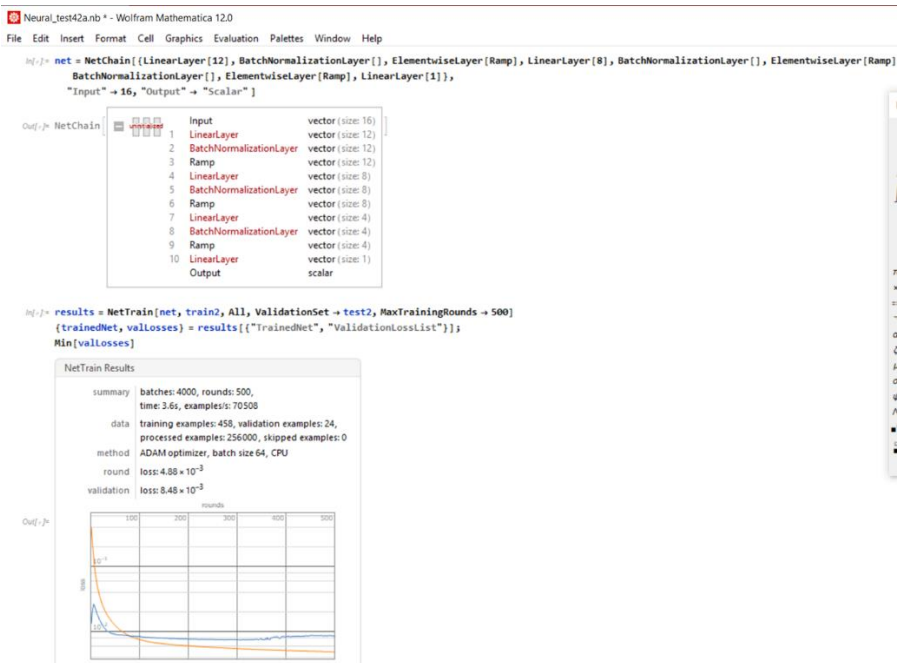

Fig. 5. An example of a 10-layer network. The input of the network indicates 16-dimensional vector—the size of state space of the price changes and the scalar output—training on, or prediction of one of the three monthly indexes.

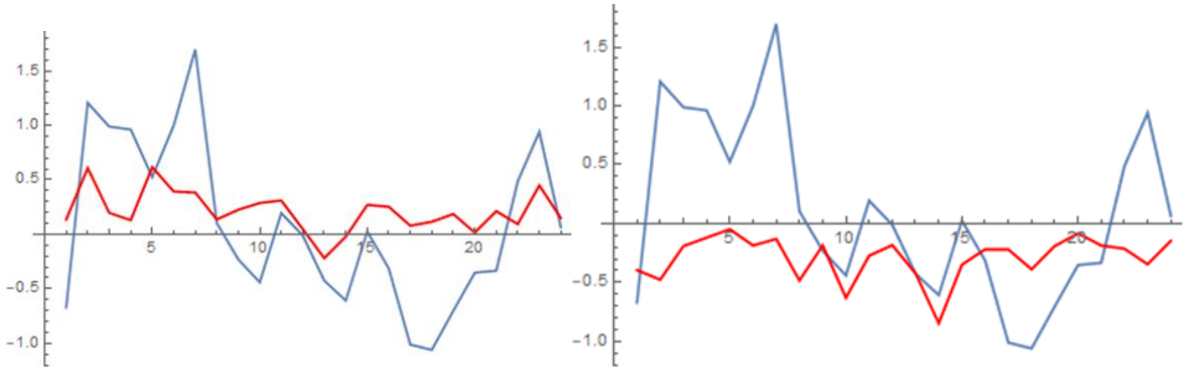

Fig. 6. Examples of the Chinese sentiment index (blue) and its backward prediction (red) by a 10-layer scalar network from buy quotes (correlation ρ=0.5841) and sell quotes (correlation ρ=0.2934) of randomly selected traders. Traders are "uninformed", i.e. they make predictions based on their own data through a network trained by an informed trader. Results are poorly reproducible.

Table 4. Select runs of 10-layer neural network for the backward prediction of monthly indexes from traders' own activity. Capital latters B, S and I mean "Buy", "Sell" and "Imbalance" samples. Letters from a-d refer to a particular trader. An arrow designates training vs. prediction tapes. The symbols $r_{1-3}$ refer to the correlation coefficients of the neural network predictor and actual indexes. Explanatory power of the predictions can be inferred from the squares of correlation coefficient. For instance, if we treat broker "b" as informed, trader "a" could have predicted bond yield for the next month from her imbalances with an explanatory power $\rho = r_3^2 = 0.5178^2 \approx 26.8\%$. Test results were poorly reproducible on successive runs of the network.

| Trader/B, S, I | Sentiment, $r_1$ | Market return $r_2$ | BOC Bond, $r_3$ | Notes |
|---|---|---|---|---|
| Bb→Ia | 0.0389 | 0.4099 (-0.2888) | 0.5178 | (Tanh) instead of ReLU |
| Bb→Ib | -0.1663 | 0.2607 | -0.1339 | |
| Bb→Ic | 0.3075 | -0.3951 | -0.2713 | |
| Bb→Id | 0.4600 | 0.3936 | 0.2525 | |
| Sb→Bc | -0.0431 | -0.3944 | -0.2389 | Control |
| Ic→Bb | -0.2255 | -0.3282 | 0.1444 | Control |

Our third exercise was to use a 7-layer convolutional neural network (CNN) sketched on the Fig. 6. CNN is conventionally used for image recognition and analysis. Essentially, we used matrices of output regression (Example provided in Fig.1) as if they were digitized information for the visual images to predict direction of an index. Table 4 demonstrates trials with randomly selected informed traders in both training and predictive samples as well training samples with only "uninformed" traders, i.e. the traders who observe only bid and ask volumes per basket, without access to current or past magnitudes of the index. Unlike the results from Table 3, the statistically significant results from Table 5 were broadly reproducible on successive runs of the network.

General conclusion from the Table 5 is that CNN can reliably predict Chinese market sentiment from daily imbalances, prediction of the stock market returns is usually significant at 10% but not at 5%, and the yield of BOC bond cannot be inferred from the imbalances. The implication of insider information does not improve prediction very much for the market sentiment index, somewhat helps to predict the direction of the stock index but well within assumed 10% statistical dispersion of the results and is irrelevant for the direction of bond yields. In all cases there is little difference whether "uninformed" trader trains her network on the imbalances of her informed colleague, or another uninformed trader.

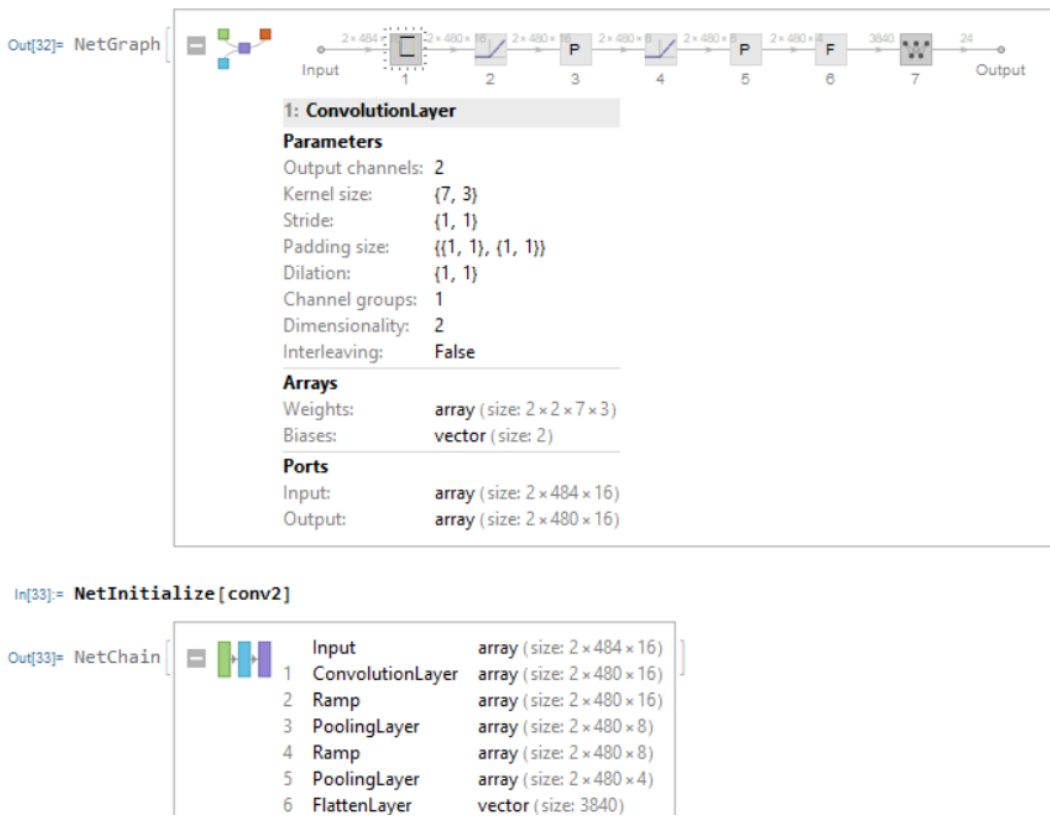

Fig. 7. Sketch of the 7-layer CNN network.

The fact that Chinese bond yields do not seem to influence orders on the Chinese stock market suggests that equity market risk dominated changes in the risk-free rate so much during the years 2009-2010—the situation typical in developing markets—that the brokerage clients ignored this information when placing their orders. Given the results of our three tests, we use 7-layer CNN network—essentially treating the matrix obtained from the price-at-volume time series—as a picture, as the most reliable.

Table 5. Correlation of predictions of the three indexes—Chinese market sentiment, returns of Shanghai stock index or 10-year bellwether bond of Bank of China—from six runs of the 7-layer CNN. The baseline batch length is 50 but we experimented with longer runs (500, 1500) and it rarely improves the precision. Traders "a" and "b" were designated as informed about the index. First two columns include informed traders in both training and prediction sample. Second pair of columns includes informed trader only in the training sample. Third pair of columns does not have an informed trader at all. Panel A) Prediction of Chinese Market Sentiment. Panel B) Prediction of the returns on the Shanghai stock market. We include results from twelve runs of the CNN, which are not significantly different. C) Prediction of the monthly yield of BOC 10-year bond. For comparisons we show the results obtained by a single 500 round and by changing ReLU into TANH perceptron function. Lower left rectangle indicates data obtained from six TANH trials. General conclusion from the table is that CNN can reliably predict Chinese market sentiment from daily imbalances, prediction of the stock market returns is significant at 10% but not at 5%, and the yield of BOC bond is uncorrelated with the imbalances.

A)

| Predicted index | Sentiment | | | | | |
|---|---|---|---|---|---|---|
| Traders training-predict | a-c | b-c | b-c | c-d | c-d | d-e |
| | 0.9756 | 0.9382 | 0.9505 | 0.9531 | 0.9120 | 0.9166 |
| | 0.9527 | 0.9643 | 0.9474 | 0.9669 | 0.9346 | 0.9014 |
| | 0.9545 | 0.9007 | 0.9232 | 0.9399 | 0.9455 | 0.9307 |
| | 0.9684 | 0.9518 | 0.9483 | 0.9695 | 0.9569 | 0.9660 |
| | 0.9727 | 0.9364 | 0.9676 | 0.9699 | 0.9344 | 0.9431 |
| | 0.9697 | 0.9371 | 0.9476 | 0.9636 | 0.9581 | 0.9225 |
| **Average** | 0.9656 | 0.9381 | 0.9475 | 0.9605 | 0.9403 | 0.9300 |
| **Student Var at 10%** | 0.0139 | 0.0307 | 0.02038 | 0.0170 | 0.0249 | 0.0324 |

B)

| Predicted index | Stock index return | | | | | |
|---|---|---|---|---|---|---|
| | a-c | b-c | b-c | c-d | c-d | d-e |
| | 0.4169 | 0.4056 | 0.6882 | 0.7065 | 0.1820 | 0.2083 |
| | 0.5363 | 0.3775 | 0.4370 | 0.4296 | 0.5341 | 0.1820 |
| | 0.1017 | -0.1194 | 0.2451 | 0.2789 | 0.4510 | 0.1656 |
| | 0.0537 | 0.3293 | 0.2767 | 0.4279 | 0.2528 | 0.0906 |
| | 0.3947 | 0.4636 | 0.6770 | 0.4051 | 0.0693 | 0.4107 |
| | 0.4310 | 0.3038 | 0.3851 | 0.3717 | 0.3207 | 0.6272 |
| | | | | | | |
| **Average-6** | 0.3223 | 0.2933 | 0.4515 | 0.4366 | 0.3017 | 0.2807 |
| **Student Var at 10%** | 0.2826 | 0.3023 | 0.2766 | 0.2067 | 0.2472 | 0.2889 |
| **Average-12** | 0.3385 | 0.2710 | | | | |
| **Student Var at 10%** | 0.2280 | 0.2374 | | | | |

C)

| Predicted index | Yield on 10-year BOC bond | | | | | |
|---|---|---|---|---|---|---|
| | a-c | b-c | b-c | c-d | c-d | d-e |
| | 0.0020 | -0.1425 | -0.0575 | -0.0281 | -0.1188 | -0.0296 |
| | 0.0413 | -0.1958 | -0.0556 | 0.1921 | 0.1953 | 0.0502 |
| | -0.0729 | -0.1434 | -0.2029 | -0.1040 | 0.1149 | -0.1185 |
| | 0.0019 | 0.0006 | 0.3416 | 0.0620 | -0.0514 | 0.1155 |
| | 0.1427 | 0.2798 | -0.0575 | -0.170 | -0.3020 | -0.4854 |
| | -0.5078 | -0.3818 | 0.1061 | 0.0081 | -0.1738 | -0.0613 |
| | | | | | | |
| **Average** | 0.1766 | 0.1078 | 0.0546 | 0.1236 | 0.1079 | 0.1032 |
| **Student Var at 10%** | 0.1819 | 0.2990 | 0.2623 | 0.1793 | 0.2700 | 0.2260 |

|  |  |  |  |  | -0.1407 | -0.2493 |
|---|---|---|---|---|---|---|
| **500 Rounds** | 0.1415 | 0.2063 | -0.2305 | 0.1633 |  |  |
| **TANH** | -0.0517 | 0.3209 | -0.2302 | 0.0754 | 0.02279 | -0.0748 |
| **Student Var at 10%, TANH** |  |  |  |  | 0.2755 | 0.3273 |

## 7. Empirical liquidity of the Chinese stock market in the period 2009-2010

A proposed microstructure model of the Chinese stock market allows us to analyze both predictable and unpredictable frictions, resulting from two interleaving factors: 1) imperfect balance between buy and sell orders and 2) securities changing value during trading.

The net cost of trading is computed similarly to Menkveld, 2017, though their formula can accept different conventions. Our formula presumes that the brokerage sells asset in today's quantity marked to market at yesterday's buy price and replenishes its inventory sold yesterday at today's ask price and the cost to the customer is the same in value and opposite in magnitude. Of course, the signs in Equation (8) are arbitrary.

$$\pi_{ti} = p_{ai,t-1}V_{b,t} - p_{bi,t-1}V_{s,t} \tag{8}$$

In Equation (8), $\pi_t$ is our definition of the cost of trading, and $p_a$, $p_b$—are the ask/bid price buckets. The $V_b$, $V_s$—are the volumes of buy/sell orders. The index $i=1$-$16$ signifies the price bucket. Note that in market equilibrium, in the Equation (7) cost averaged over all buckets is equal to the (constant) bid-ask spread times the daily turnover and is always non-negative. Outside of equilibrium, the sign of $\pi_t$ can be arbitrary because of fluctuating stock prices.

Our analysis by the CNN indicates that the net cost of trading is a fair predictor of the market direction in the sense that we have outlined in a previous section (Fig. 7). Namely, if the broker or regulator is "blind" with respect to the order size, she can get a clear idea about the Chinese market sentiment from the trading costs only. Her idea of stock market direction would be imperfect but statistically significant and, finally, there is no connection to the Bank of China bond prices through our model.

While this exercise is purely imaginary with respect to the Chinese brokerages, we suggest that this conclusion—that, in the observed period, trading costs reflected market sentiment more or less mechanically—can help traders and regulators alike in the case of the "Dark Pools". In latter case, the information about exchange's own strategy is covert and can be gleaned only indirectly.

The Equation (8) can be recast in the (dynamic) version of Amihud (1986) liquidity measure. Namely,

$$\lambda_{ti} = \frac{|\pi_{ti}|}{(V_{b,t}+V_{a,t-1})/2} \tag{9}$$

Further on, we use liquidity lambda to predict the same indexes. Intuitive meaning of our version of the Amihud measure is that it represents average cost for the agent to make a roundtrip inside the same price bucket with one share. To provide a glimpse of the magnitude and volatility of λ, we display its daily dynamics in the Fig. 8. Our sample includes a day of the Flash Crash in US stock markets (May 6, 2010), which can be, dependent on daytime, either trading day 326 or 327 in our sample. There is no visible anomaly of the liquidity of the Chinese stock market during or after that day.

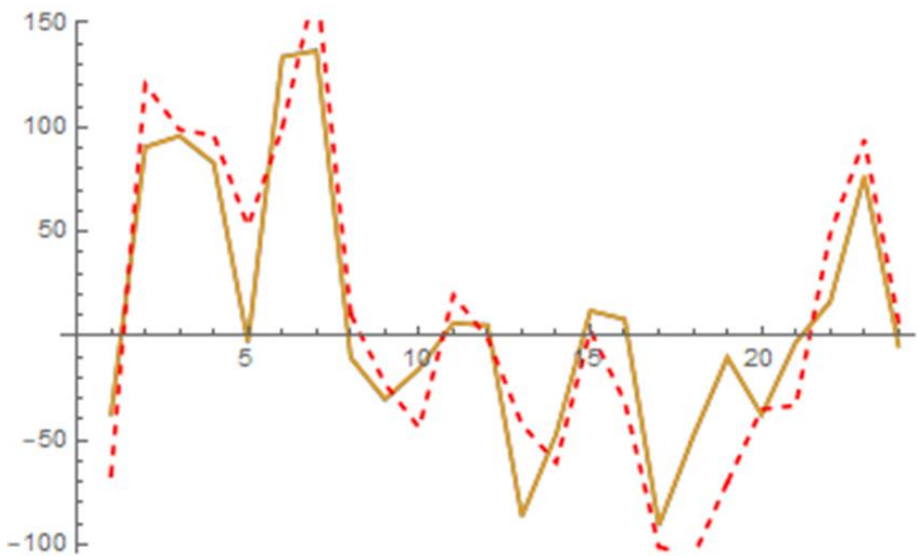

Fig. 7. Correlation of CNN predictions of the market sentiment from the unexpected monthly cost-of-trading (PC, brown, solid) with predictions unexpected order volume (MS, red, dash). $\rho_{\varepsilon,PC} \approx \rho_{\varepsilon,MC} \approx 0.913$. For comparability, the scale of market sentiment was increased by 100.

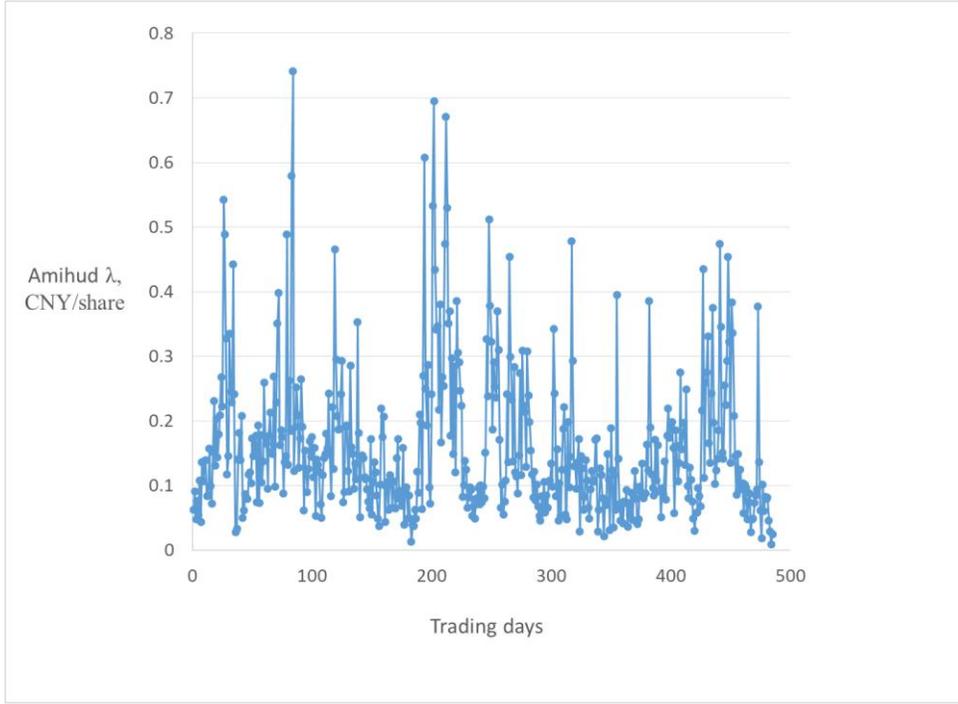

Fig. 8. Time series of average Amihud λ (Equation 9) across all price buckets for a single trader.

Our naïve visual intuition is supported by the CNN analysis of the regression residuals. We display the testing strategy in the Fig. 9. The observation period (years 2009-2010) is split into 8 samples of 60 days each. One sample of two adjacent periods (usually, but not necessarily the first) is used for network training. This constitutes one training and five predictor samples. We then try to predict monthly indexes backwards from the training data. Null hypothesis is formulated as follows:

*H0: For i=1÷5, Correlation[Index prediction from $\lambda_i$, Index]=Correlation[Average[λ] ,Index]*

*H1: For at least one i, Correlation[Index prediction from $\lambda_i$, Index] ≠ Correlation[Average[λ] ,Index]*

Intuitive meaning of the null hypothesis above is that predictions obtained from subsamples of illiquidity measure are no different from each other. We, of course, would prefer if the null can be rejected for the samples, containing the Flash Crash in America. Statistical significance of the correlations of the predicted monthly indexes for one of the traders (tapes) is shown in Table 5.

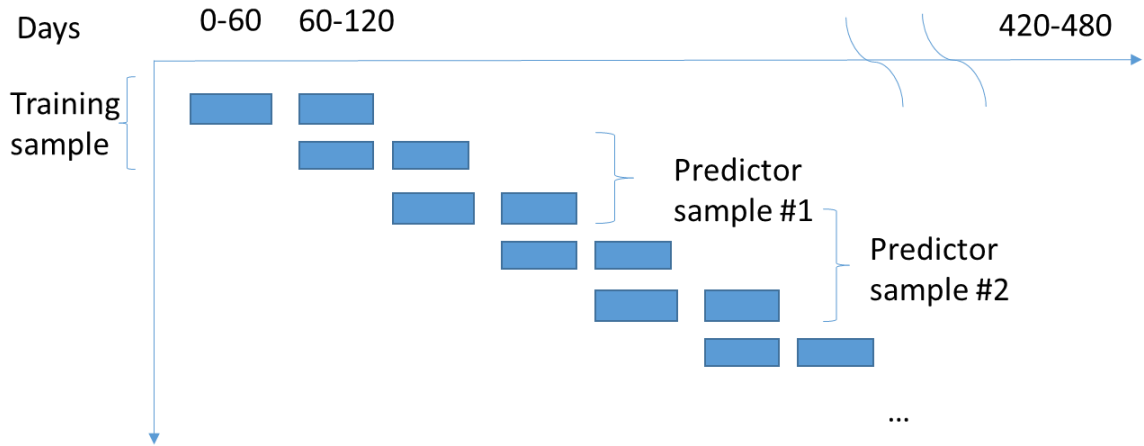

Fig. 9 Schematic description of testing procedure. We use the data on λ, which we consider a measure of market friction, from the training sample to predict indexes from λ's in five other samples. In the drawing above, the training sample comes first in calendar time but its position within entire sample can be arbitrary.

Table 5. Probabilities of null hypothesis H0 from a typical correlation test of index predictions from Amihud λ. As a rule, null cannot be rejected for any of the three indexes. Note that in agreement with the results of Section 6, almost no statistically significant correlation of prediction of the return on the stock index and none—for the yields on the 10-year bond can be observed as well. On the contrary, prediction of the market sentiment from Amihud illiquidity measure is robust. This corresponds to H0, H1 above where the right hand side is replaced by zero. A) Probabilities for prediction of sentiment index, B) probabilities for prediction of stock market returns, C) probabilities for prediction of bond yields. Highlighted are the subsamples, which include the US Flash Crash of May, 6, 2010.

A)

| Trading days in sample | P from Pearson | P from Spearman |
|---|---|---|
| 120-240 | 0.364 | 0.932 |
| 180-300 | 0.605 | 0.976 |
| 240-360 | 0.227 | 0.062 |
| 300-420 | 0.847 | 0.342 |
| 360-480 | 0.975 | 0.903 |

B)

| Trading days in sample | P from Pearson | P from Spearman |
|---|---|---|
| 120-240 | 0.400 | 0.712 |
| 180-300 | NA | 0.561 |
| 240-360 | 0.452 | 0.437 |
| 300-420 | 0.857 | 0.806 |
| 360-480 | 0.616 | 0.061 |

C)

| Trading days in sample | P from Pearson | P from Spearman |
|---|---|---|
| 120-240 | 0.622 | 0.285 |
| 180-300 | 0.721 | 0.565 |
| 240-360 | 0.926 | 0.743 |
| 300-420 | 0.688 | 0.554 |
| 360-480 | 0.167 | 0.143 |

We observe from the Table 5 null hypothesis, namely, that illiquidity in subsamples that include or exclude the date of American Flash Crash behaves any different from other samples, cannot be rejected. Only one probability in the Table 5 is below 10% and it is not stable with respect to the consecutive runs of the neural network with a different seed.

## 8. Conclusion

In our paper, we propose a microstructure model of the Chinese stock market. We build it from a state space of interday correlations of volumes between price buckets. For the 100%-efficient market in equilibrium, our time series would be exactly equal to zero.

This model is analyzed by OLS in a *dual* state space, connected to an original state space by a Fourier transform across the multiple price buckets. This procedure corresponds to an approximation of the time evolution by a pseudodifferential operator of a general form, which, in discrete case can be represented by a matrix of arbitrary dimension (selected by convenience; 32×32 in our case) acting on the space of the Fourier coefficients (see Appendix for details). This method is completely general and can be applied to any time series, which can be grouped into panels—volumes per a given stock price in our case.

The main conclusion of our analysis is that *unpredictable* dynamics of trades completed by the Chinese brokers in the period 2009-2010 was tightly correlated with the Chinese market sentiment in a highly nonlinear sense of machine learning. This can be interpreted as retail investors trading according to the available market information they receive, following a herd mentality. Yet, the returns on the stock index are only partly predictable with the probability of the null—that their prediction is totally random—not much below of 10%. We observed no connection between the trading activity and the yields on bellwether ten-year bond of the Bank of China. This may signify that the market risk in emerging markets, which was the Chinese stock market in 2009-2010, has only a small dependence on prevailing borrowing rates.

Finally, we tested whether the Flash Crash of American markets on May 6, 2010, was reflected in the liquidity of the Chinese stock market. For that we used a dynamic version of Amihud illiquidity measure $\lambda$. So far, we did not find any evidence that liquidity was higher or lower than the average during the period preceding or following the Flash Crash. This indicates a need for a finer measure of contagion between American and Chinese stock markets.

**Appendix**. Pseudodifferential operators

Frequently financial time series are described by the *AR(n)* models, which are, in their turn, are the discrete analogues of the differential operators with constant coefficients. Pseudodifferential operators can be considered as generalizations of *ARMA(p,q)* models. From that angle, conventional *ARMA(p,q)* models are discrete pseudodifferential operators with a rational function as a symbol. Formal definition of the pseudodifferential operator goes as follows (e.g. Taylor, 1981, Hörmander, 1987)

$$A(u) = \frac{1}{(2\pi)^n} \int_{\mathbb{R}^{2n}} e^{-i(x-y)\xi} a(x,y,\xi) u(y) dy d\xi \tag{A.1}$$

One can easily write a solution for a Cauchy problem for Kolmogorov-Fokker equation, describing Ito diffusion through a pseudodifferential operator. Indeed,

$$\frac{\partial \psi}{\partial t} = \mathcal{L}(\psi)$$

$$\psi(x,0) = f(x) \tag{A.2}$$

where $\mathcal{L}$ is a generator of Ito diffusion of the following form. Here, a is a n-dimensional vector, $\Sigma$ is n×n matrix and i,j=1÷n—dimensions of the state space.

$$\mathcal{L} = \vec{a} \cdot \partial_i + \partial_i \hat{\Sigma} \partial_j \tag{A.3}$$

This problem's solution expressed in a form of Equation (A.2) looks as:

$$\psi(x,t) = \frac{1}{(2\pi)^n} \int_{\mathbb{R}^{2n}} e^{ik(x-y)} A(k,t) \cdot f(y) d^n y d^n k \tag{A.4}$$

With

$$A(k,t) = e^{(i\vec{k}\cdot\vec{a} - \vec{k}^T \hat{\Sigma} \vec{k})t} \tag{A.5}$$

In our case, the phase-space regression has a form of Equation (A.4) of the main text:

$$\vec{X}_{t+1,\omega'} - \vec{X}_{t,\omega'} = \hat{\beta}_{\omega'\omega} \vec{X}_{t,\omega} + \vec{e}_{t+1,\omega} \delta_{\omega\omega'} \tag{A.6}$$

To better demonstrate connection of this regression to the pseudodifferential operators, we replace discrete time steps in Equation (A.6) with continuous time. The transfer Equation (A.6) becomes:

$$d\vec{X}(t)_{\omega'} = \hat{\beta}_{\omega'\omega}\vec{X}(t)_\omega + d\vec{\varepsilon}(t)_\omega \delta_{\omega\omega'}$$

The value of the state vector as a function of a state variable *x* for an arbitrary time *T*, can be expressed as a moving average-type equation:

$$\vec{X}(x,T) = \left(\int e^{\beta_{\omega'\omega}(T-t)+i(\omega'-\omega)t} dt\right) \vec{X}(x,0) + \int \sum_{\omega,\omega'} e^{\beta_{\omega'\omega}(T-t)+i(\omega'-\omega)t} \vec{\varepsilon}(t)_\omega dt \qquad (A.7)$$

In the Equation (A.4), the symbol of the pseudodifferential operator is equal to:

$$A(\omega,t,T) = \sum_{\omega,\omega'} e^{\beta_{\omega'\omega}(T-t)} \qquad (A.8)$$